\documentclass{moriond}

\usepackage{amsmath,amssymb,cite}

\newcommand{\GeV}{\,\text{GeV}}
\newcommand{\TeV}{\,\text{TeV}}

\newcommand{\beq}{\begin{equation}}
\newcommand{\eeq}{\end{equation}}
\renewcommand{\Re}{\text{Re}\,}

	\newcommand{\Photo}{\includegraphics[height=35mm]{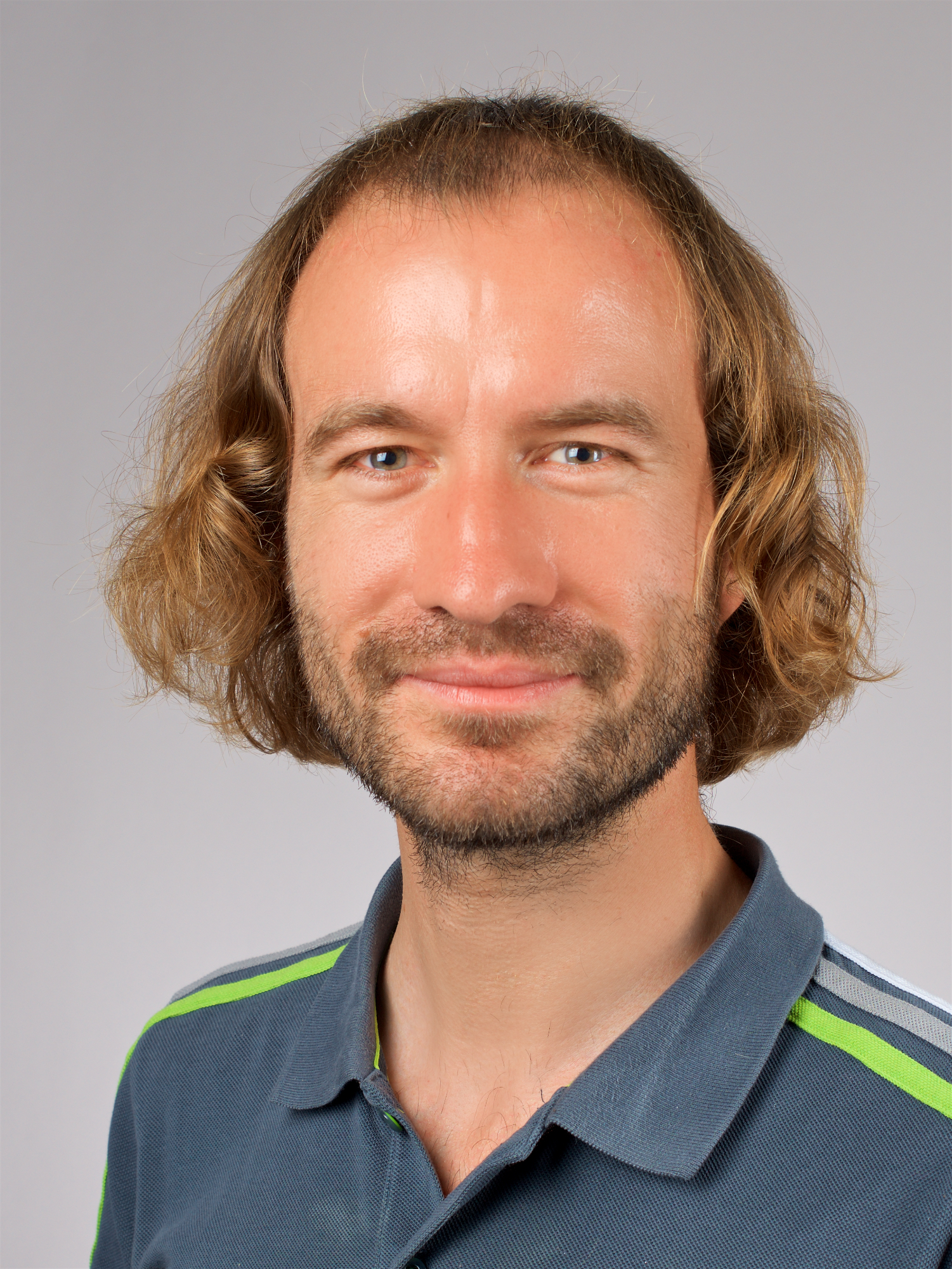}}
\begin{document}

\vspace*{4cm}
\title{EXPLAINING THE CABIBBO ANGLE ANOMALY}

\author{A.\ CRIVELLIN$^{1,2}$}

\address{$^1$Paul Scherrer Institut, CH--5232 Villigen PSI, Switzerland\\
$^2$Physik-Institut, Universit\"at Z\"urich, Winterthurerstrasse 190, CH--8057 Z\"urich, Switzerland}

\maketitle
\abstracts{The determinations of the Cabibbo angle from kaon, pion, $D$ and tau decays disagree with the one from super-allowed beta decays. The resulting $\approx 3\,\sigma$ deficit in first row (but also first column) CKM unitarity is known as the Cabibbo angle anomaly (CAA). Determining $V_{ud}$ from beta decays requires knowledge of the Fermi constant $G_F$, usually extracted from muon decay. However, because new physics might also affect muon decay, and thus the determination of the Fermi constant, an interesting interplay with the global electroweak fit arises where $G_F$ is a crucial input. In these proceedings we review the CAA and its different interpretations in terms of physics beyond the SM, first discussing the corresponding effect in the effective field theory and then using (simplified) models.}

\section{Introduction}

\begin{figure}[t]
	\includegraphics[width=0.55\linewidth]{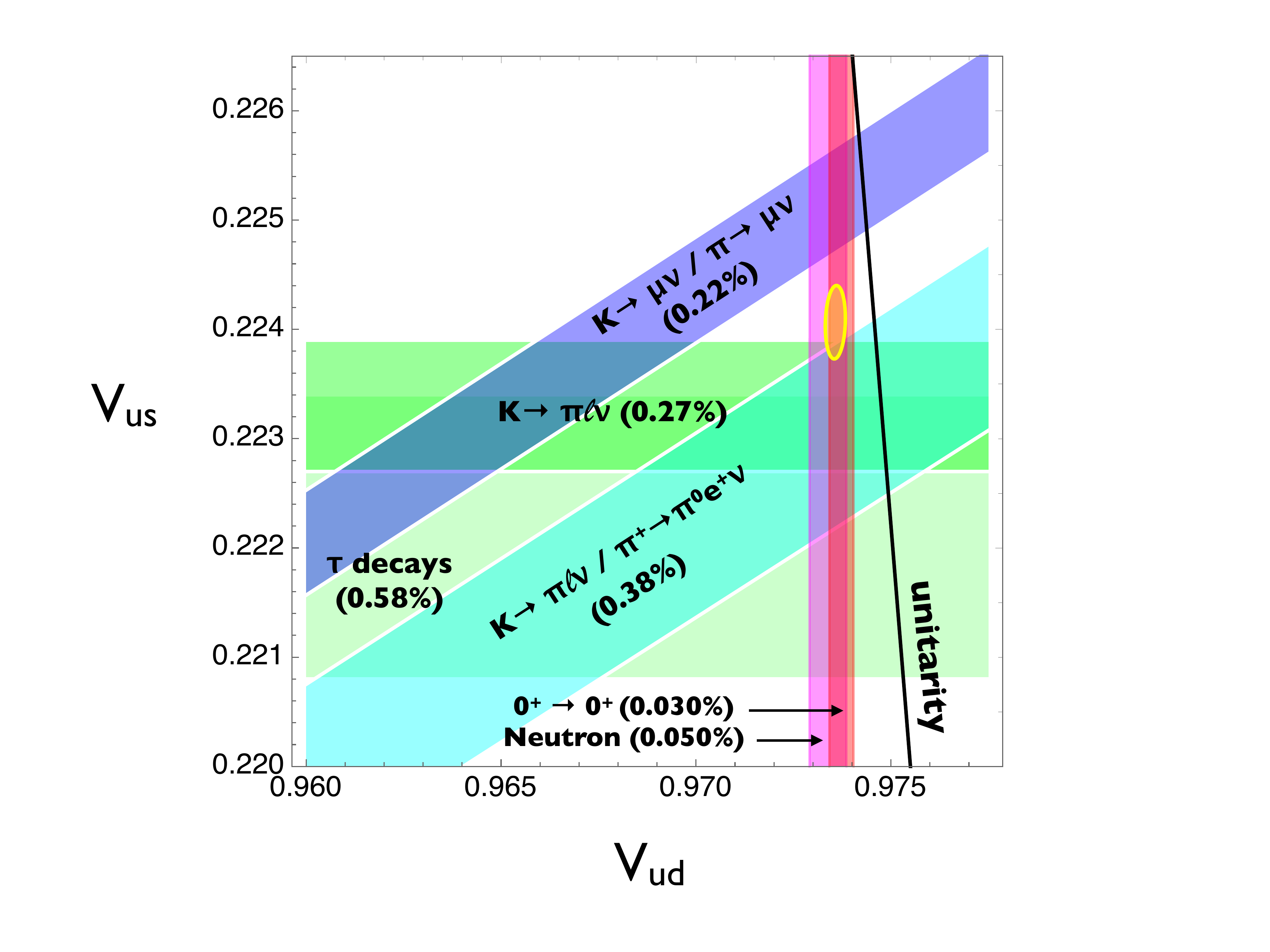}\hspace{-12mm}	\includegraphics[width=0.5\linewidth]{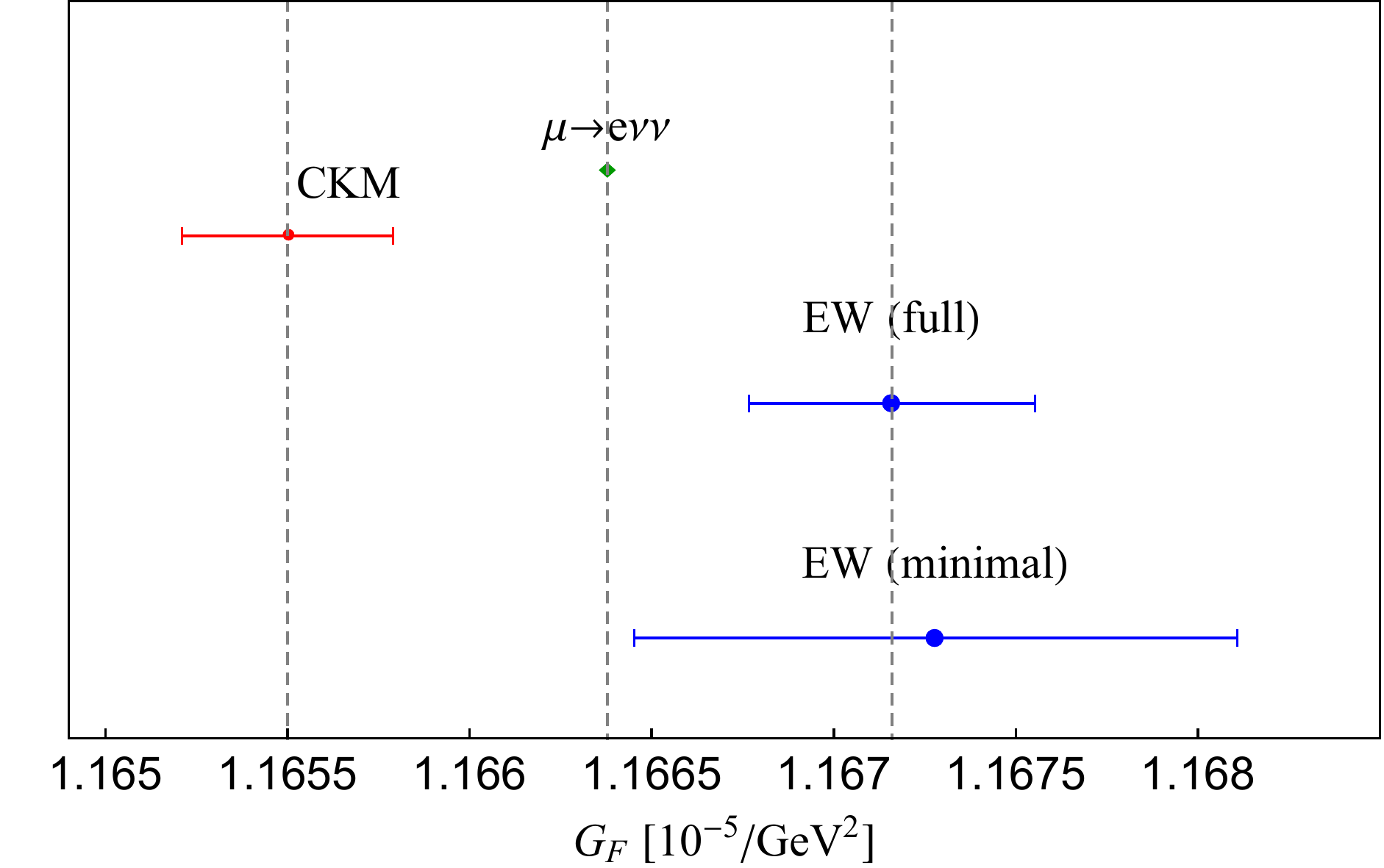}
	\caption{Left: Different determinations of $V_{us}$ and $V_{ud}$ leading to the CAA~\cite{Bryman:2021teu}. Right: Values of $G_F$ from the different determinations~\cite{Crivellin:2021njn}.\label{GFplot} }
\end{figure}

The Cabibbo angle parametrizes, up to higher order corrections in the Wolfenstein parameter, the mixing among the first two generations of quarks and therefore determines the CKM elements $V_{us}$ and $V_{ud}$. It can be measured in kaon, pion, tau and beta decays. Concerning the latter, super-allowed beta decays provide the most precise determination of the Cabibbo angle, however, its values disagrees with the other determinations. The significance of this tension crucially depends on the radiative corrections applied to $\beta$ decays~\cite{Hardy:2020qwl}, but also on the treatment of tensions between $K_{\ell 2}$ and $K_{\ell 3}$ 
decays~\cite{Moulson:2017ive} and the constraints from $\tau$ decays~\cite{Amhis:2019ckw}, see Ref.~\cite{Crivellin:2020lzu} for more details. However, in the end, an $\approx 3\,\sigma$ deficit in first row (and less significant in first column) CKM unitarity~\cite{Zyla:2020zbs}
\begin{align}
\big|V_{ud}\big|^2+\big|V_{us}\big|^2+\big|V_{ub}\big|^2
= 0.9985(5)\,,\qquad\big|V_{ud}\big|^2+\big|V_{cd}\big|^2+\big|V_{td}\big|^2 = 0.9970(18)\,,
\label{1throw}
\end{align}
seems reasonable. This situation is shown in the left plot of Fig.~\ref{GFplot}~\cite{Bryman:2021teu}.

As these unitarity relations are dominated by $V_{ud}$, and since tau, pion and kaon and $D$ decays would require different kinds of new physics contributions, this suggests that NP should enter the determination of $V_{ud}$, assuming that the anomaly has, in fact, a beyond the Standard Model (BSM) origin. Furthermore, as shown in Ref.~\cite{Crivellin:2020lzu}, the sensitivity to a BSM effect in (super-allowed) $\beta$ decays is enhanced by a factor $|V_{ud}|^2/|V_{us}|^2$ compared to kaon, $\tau$, or $D$ decays.

Importantly, for determining the Cabibbo angle from (super-allowed) beta decays, knowledge of the Fermi constant $G_F$, i.e.~the effective weak force at low energies, is necessary. Even though $G_F$ is extremely precisely measured via muon decay~\cite{Gorringe:2015cma}
\begin{equation}
\label{Mulan}
G_F^\mu=1.1663787(6) \times 10^{-5} \GeV^{-2}\,,
\end{equation}
at the level of $0.5\,\text{ppm}$, it is not necessarily free of BSM contributions. In fact, as for any measurement, one can only conclude the presence or absence of BSM effects by comparing $G_F^\mu$ to another independent determination. Therefore, the CAA could not only be explained via a BSM effect in beta decays, i.e.~$d\to u e\nu$ transitions, but also via an effect in muon decay ($\mu\to e\nu\nu$), changing the value of the Fermi constant. 

In order to determine if muon decay and/or beta decays subsume BSM effects, one can determine the Fermi constant from the EW precision observables~\cite{Marciano:1999ih}. In our global fit~\cite{Crivellin:2021njn}, including the $W$ mass, $\sin^2\theta_W$, and $Z$-pole observables~\cite{Schael:2013ita} etc., we find 
\begin{equation}
\label{GFEW}
G_F^\text{EW}\Big|_\text{full}=1.16716(39)\times 10^{-5}\GeV^{-2}\,.
\end{equation}
As depicted in the right plot of Fig.~\ref{GFplot}, this value lies above $G_F^\mu$ by $\approx2\sigma$, reflecting the known tensions within the EW fit~\cite{Baak:2014ora,deBlas:2016ojx}. For comparison with Ref.~\cite{Marciano:1999ih}, if one considered only $\sin^2\theta_W$, one would obtain
\begin{equation}
G_F^\text{EW}\Big|_\text{minimal}=1.16728(83)\times 10^{-5}\GeV^{-2},
\end{equation}
consistent with Eq.~\eqref{GFEW}, but with a larger uncertainty. The pull of $G_F^\text{EW}$ away from $G_F^\mu$ is mainly driven by $M_W$, $\sin^2\theta_W$ from the hadron colliders, $A_\ell$, and $A_{\rm FB}^{0, \ell}$.

We can also translate the CAA into a determination of the Fermi constant by requiring that it takes the value which restores CKM unitarity:
\begin{equation}
\label{GFCKM}
G_F^\text{CKM} = 0.99925(25)\times G_F^\mu =1.16550(29)\times 10^{-5}\GeV^{-2}\,.
\end{equation}
Comparing the three independent determinations of $G_F$ in right plot of Fig.~\ref{GFplot}, one finds a situation in which $G_F^\text{EW}$ lies above $G_F^\mu$ by $2\sigma$,  $G_F^\text{CKM}$ below $G_F^\mu$ by $3\sigma$, and the tension between $G_F^\text{EW}$ and $G_F^\text{CKM}$ amounts to $3.4\sigma$. Therefore, if one wants to explain the CAA, i.e.~the disagreement between $G_F^{\rm CKM}$ and $G_F^{\mu}$ by NP in muon decay only, one increases the tension with the EW. Therefore, models that give a direct effect in beta decays and/or an effect in the EW fit are welcome. 

In these proceedings, we first review the possible solutions to the CAA in terms of effective operators before we briefly discuss (simplified) new physics models.

\begin{figure}[t!]
	\centering
	\includegraphics[width=0.7\textwidth]{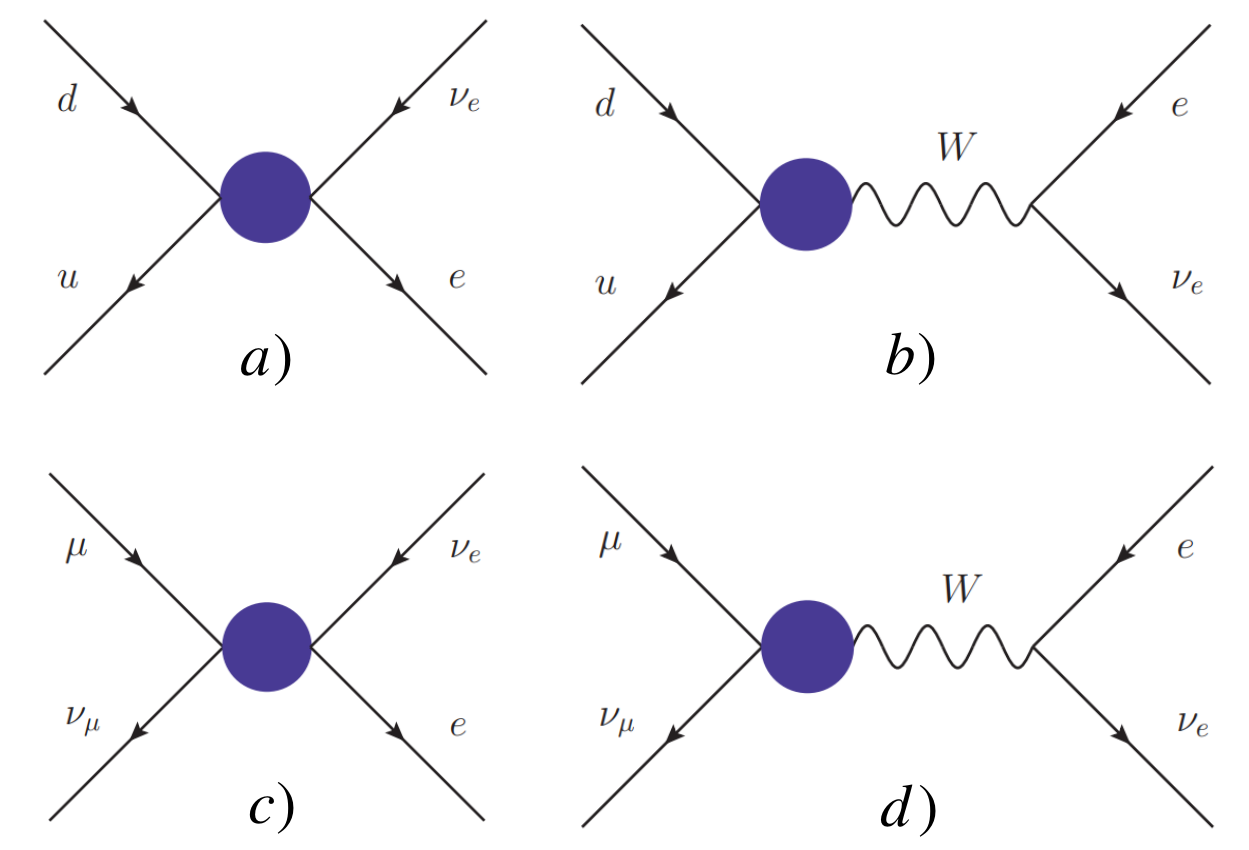} 
	\hspace{10mm}
	\caption{Possible explanations of the CAA in terms of effective operators. a) and b) change directly the determination of $V_{ud}$ from beta decays while c) and d) lead to an effect in the Fermi constant that changes indirectly the values of $V_{ud}$ extracted from (super-allowed) beta decays.
		\label{CAAexplanations}	}
\end{figure}

\section{SMEFT analysis}

Among the set of gauge-invariant dimension-$6$ operators~\cite{Buchmuller:1985jz,Grzadkowski:2010es}, there are four classes that affect muon and/or beta decays~\cite{Han:2004az,Falkowski:2014tna,Ellis:2018gqa,Skiba:2020msb} and therefore have the potential to explain the CAA:
\begin{itemize}
	\vspace{-1mm}
	\item[a)] 4-fermion operators affecting beta decays ($u\to d e\nu$)
	\vspace{-1mm}
	\item[b)] Operators generating modified $W$--$u$--$d$ coupling,
	\vspace{-1mm}
	\item[c)] 4-fermion operators affecting muon decays ($\mu\to e\nu\nu$)
	\vspace{-1mm}
	\item[d)] Operators generating modified $W$--$\mu$--$\nu$ couplings
	\vspace{-1mm}
\end{itemize}
This is illustrated in Fig.~\ref{CAAexplanations}.

\subsection{Four-fermion operators in ${d\to u e\nu}$}

The operators $Q_{\ell equ}^{\left( 1 \right)1111}$ and $Q_{\ell equ}^{\left( 3 \right)1111}$ give rise to $d\to u e\nu$ scalar amplitudes which lead to enhanced effects in $\pi\to\mu\nu/\pi\to e\nu$ (w.r.t~$\beta$ decays) and therefore can only have a negligible impact on $V_{ud}$. Furthermore, the tensor amplitude generated by $Q_{\ell equ}^{\left( 3 \right)ijkl}$ has a suppressed matrix element in $\beta$ decays. 

\begin{figure}[t!]
	\centering
	\includegraphics[width=0.7\textwidth]{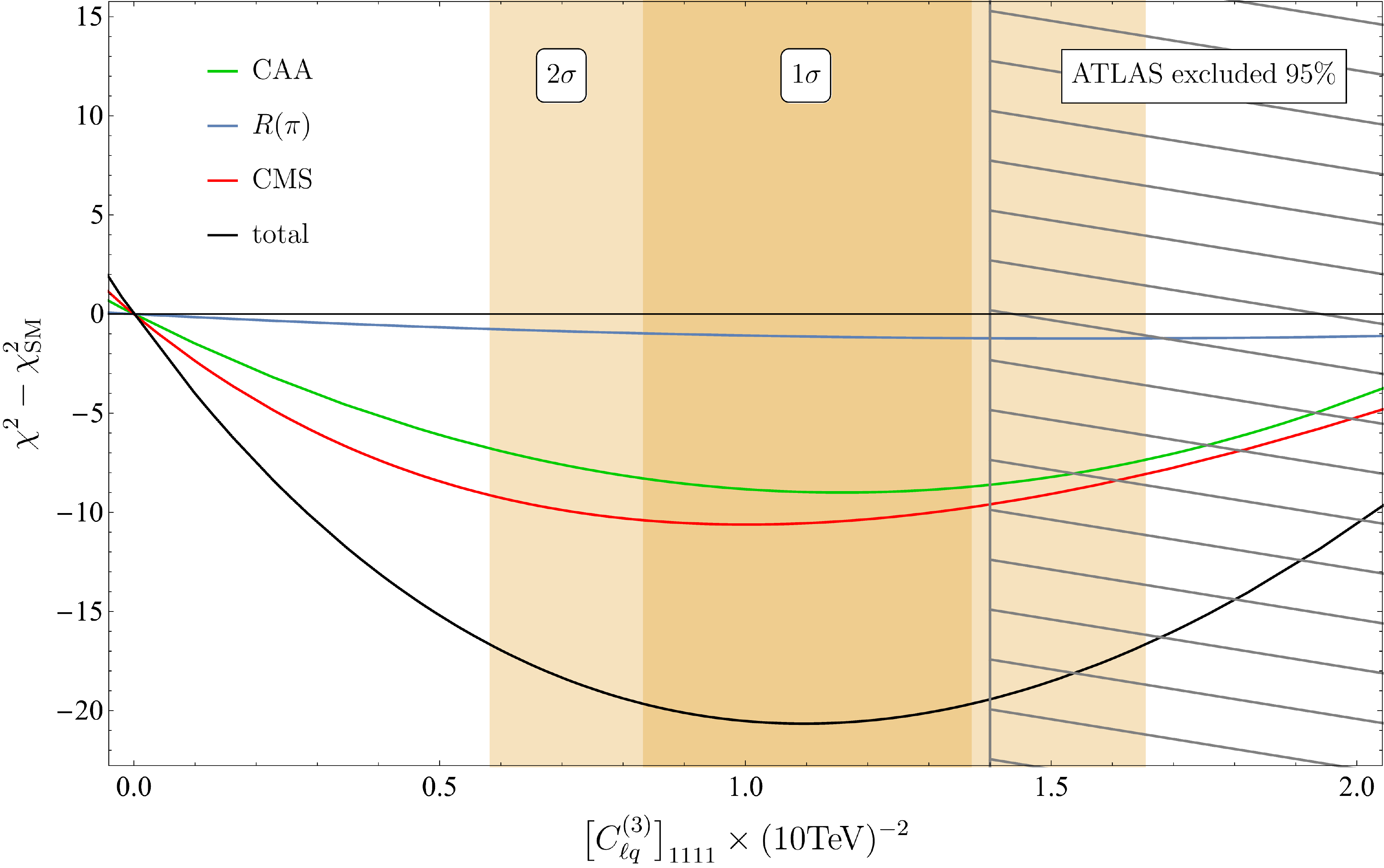} 
	\hspace{10mm}
	\caption{
		$\Delta \chi^2 = \chi^2 - \chi^2_{SM}$ as a function of the Wilson coefficient $[C_{\ell q}^{(3)}]_{1111}$ for $R(\pi)$ (blue), the CAA (green), the CMS analysis of di-lepton pairs  (red) and the combination of them (black) with the dark orange region showing the $1 \sigma$ and $2 \sigma$ regions of the combined fit. The hatched region is excluded at 95\% CL from the non-resonant di-lepton search of ATLAS (not included in the $\chi^2$ function). The best fit point for the combined fit (black) is at $[C_{\ell q}^{(3)}]_{1111}\approx1.1/(10\,{\rm TeV})^2$ where $\chi^2-\chi^2_{SM} \approx - 20$.  \label{chi2plot}	}
\end{figure}

For $Q_{\ell q}^{\left( 3 \right)1111}$, the CAA prefers
$C_{\ell q}^{\left( 3 \right)1111}\approx 0.7\times 10^{-3}G_F$.
Via $SU(2)_L$ invariance, this operator generates neutral currents 
\begin{equation}
{{\cal L}_{{\rm{NC}}}} = C_{\ell q}^{\left( 3 \right)1111}\left( {\bar d{\gamma ^\mu }{P_L}d - \bar u{\gamma ^\mu }{P_L}u} \right)\bar e{\gamma _\mu }{P_L}e,
\end{equation}
leading to effects in non-resonant di-electron searches at the LHC. Interestingly, CMS found an excess of such high-energetic electrons~\cite{Sirunyan:2021khd} and also ALTAS observed more events than expected~\cite{Aad:2020otl}, as predicted by an explanation of the CAA via this operator~\cite{Crivellin:2021rbf}. This can be seen Fig.~\ref{chi2plot} where also $R(\pi)$, testable at the future PIONEER experiment~\cite{PIONEER:2022yag}, is shown.

\subsection{Modified ${W}-{u}-{d}$ couplings}

There are only two operators modifying $W$ couplings to quarks
\begin{align}
Q_{\phi q}^{\left( 3 \right)ij} &= {\phi ^\dag }i
\overset{\leftrightarrow}{D}^I_\mu
\phi {{\bar q}_i}{\gamma ^\mu }{\tau ^I}{q_j},\notag\\
Q_{\phi ud}^{ij} &= {\phi ^\dag }i\overset{\leftrightarrow}{D}_\mu \phi {{\bar u}_i}{\gamma ^\mu }{d_j}.
\end{align}
First of all, $Q_{\phi ud}^{ij}$ generates right-handed $W$--quark couplings. $Q_{\phi q}^{\left( 3 \right)ij}$ generates modifications of the left-handed $W$--quark couplings and data prefer
\begin{equation}
C_{\phi q}^{\left( 3 \right)11} \approx 0.7\times 10^{-3}G_F.
\end{equation}
Due to $SU(2)_L$ invariance, in general effects in $D^0$--$\bar D^0$ and $K^0$--$\bar K^0$ mixing are generated. However, in case of alignment with the down-sector, the effect in $D^0$--$\bar D^0$ is smaller than the experimental value and thus not constraining as the SM prediction cannot be reliably calculated. 

\subsection{Four-fermion operators in ${\mu\to e\nu\nu}$}

Disregarding flavor indices, two operators can generate a charged current involving four leptons:
\begin{align}
Q_{\ell \ell }^{ijkl} = {{\bar \ell }_i}{\gamma ^\mu }{\ell _j}{{\bar \ell }_k}{\gamma ^\mu }{\ell _l},\qquad
Q_{\ell e}^{ijkl} &= {{\bar \ell }_i}{\gamma ^\mu }{\ell _j}{{\bar e}_k}{\gamma ^\mu }{e_l}.
\end{align}
Note that for $Q_{\ell \ell }^{ijkl}$, e.g.~ the identities
$Q_{\ell \ell }^{ijkl}=Q_{\ell \ell }^{klij}=Q_{\ell \ell }^{ilkj}=Q_{\ell \ell }^{kjil}$ hold. Taking this into account, we have 9 independent operators that contribute to ${\mu\to e\nu\nu}$ at tree-level~\cite{Crivellin:2021njn}:

\begin{enumerate}
	\vspace{-1mm}
	\item $Q_{\ell \ell }^{2112}$
	gives rise to a real and SM-like amplitude. Therefore, it can give a constructive or destructive effect in the muon lifetime and does not affect the Michel parameters~\cite{Michel:1949qe}. In order explain the CAA at the $1\sigma$ level we need		$C_{\ell \ell }^{2112}\approx -1.4\times 10^{-3} G_F$.		This Wilson coefficient is constrained by LEP searches for $e^+e^-\to \mu^+\mu^-$~\cite{Schael:2013ita}
	\begin{equation}
	- \frac{4\pi}{(9.8\TeV)^2} < C_{\ell \ell }^{1221} < \frac{4\pi}{(12.2\TeV)^2},
	\end{equation}
	such that a solution of the CAA is possible.
	\vspace{-1mm}
	\item Even though $Q_{\ell e}^{2112}$ possess a vectorial Dirac structure, it leads to a scalar amplitude after using Fierz identities. Its interference with the SM is usually expressed via the Michel parameter $\eta=\Re C_{\ell e }^{2112}/(2\sqrt{2}G_F)$, leading to a correction $1-2\eta m_e/m_\mu$. The shift in $G_F^\mu$ is constrained to be at most $0.68\times 10^{-4}$~\cite{Danneberg:2005xv}, well below the required effect for solving the CAA.  
	\vspace{-1mm}	
	\item The operators $Q_{\ell \ell (e) }^{1212}$ can contribute to muon decay since the neutrino flavors are not detected. To explain the CAA we need $|C_{\ell \ell }^{1212}|\approx 0.045\,G_F$ or 	$|C_{\ell e }^{1212}|\approx 0.09\,G_F$, such that both solutions are excluded by muonium--anti-muonium oscillations ($M=\mu^+ e^-$)~\cite{Willmann:1998gd}.	
	\vspace{-1mm}		
	\item For $Q_{\ell \ell (e) }^{1112}$ again numerical values of $|C_{\ell \ell (e) }^{1112}|\approx 0.09\, G_F$ are 
	preferred (as for all the remaining Wilson coefficients in this list). Both operators give tree-level effects in $\mu\to 3 e$, e.g.,
	\begin{equation}
	\text{Br}\left[ {\mu  \to 3e} \right] = \frac{{m_\mu ^5\tau_\mu}}{{768{\pi ^3}}}{\left| {C_{\ell \ell }^{1112}} \right|^2}=0.25\bigg|\frac{C_{\ell \ell }^{1112}}{G_F}\bigg|^2,
	\end{equation}
	which exceeds the experimental limit on the branching ratio of $1.0\times 10^{-12}$~\cite{Bellgardt:1987du} by orders of magnitude (the result for $C_{\ell e }^{1112}$ is smaller by a factor $1/2$).
	\vspace{-1mm}		
	\item The operators $Q_{\ell \ell (e) }^{2212}$ and $Q_{\ell \ell (e) }^{3312}$ contribute at the one-loop level to $\mu\to e$ conversion and $\mu\to 3e$ and at the two-loop level to $\mu\to e\gamma$~\cite{Crivellin:2017rmk}. Here the current best bounds come from $\mu\to e$ conversion. Using Table~3 in Ref.~\cite{Crivellin:2017rmk} we have
	\begin{align}
	\left| {C_{\ell \ell }^{3312}} \right| <6.4\times 10^{-5} G_F,\qquad
	\left| {C_{\ell \ell }^{2212}} \right| <2.8\times 10^{-5} G_F,
	\end{align}
	again excluding a sizable BSM effect. Similarly, $Q_{\ell e }^{3312}$ and $Q_{\ell e }^{2212}$ cannot provide a viable solution of the CAA.
	\vspace{-1mm}
	\item $Q_{\ell \ell (e) }^{2312}$, $Q_{\ell \ell (e) }^{3212}$, $Q_{\ell \ell (e) }^{1312}$, and $Q_{\ell \ell (e) }^{3112}$  contribute to $\tau\to \mu \mu e$ and $\tau\to \mu ee$, respectively, which forbid a sizable effect in analogy to $\mu\to 3e$. 
	\vspace{-1mm}
\end{enumerate}

\subsection{Modified ${W}$--${\mu}$--${\nu}$ couplings}

Only the operator
\begin{align}
Q_{\phi \ell }^{(3)ij}={\phi ^\dag }i
\overset{\leftrightarrow}{D}^I_\mu
\phi {{\bar \ell}_i}{\gamma ^\mu }{\tau ^I}{\ell_j}
\end{align}
generates modified $W$--$\ell$--$\nu$ couplings at tree level. In order to avoid the stringent bounds from charged lepton flavor violation, the off-diagonal Wilson coefficients, in particular $C_{\phi \ell }^{(3)12}$, must be very small. Since they also do not generate amplitudes interfering with the SM ones, their effect can be neglected. Because $C_{\phi \ell }^{(3)11}$ affects $G_F^\mu$ and $G_F^\text{CKM}$ in the same way and thus has no effect in the CAA. Therefore, only $C_{\phi \ell }^{(3)22}>0$ can explain the CAA while $C_{\phi \ell }^{(3)22}>0$, and $|C_{\phi \ell }^{(3)22}|<|C_{\phi \ell }^{(3)11}|$ are preferred in this case by tests of lepton flavor universality tests such as $\pi(K)\to\mu\nu/\pi(K)\to e\nu$ or $\tau\to\mu\nu\nu/\tau(\mu)\to e\nu\nu$~\cite{Coutinho:2019aiy,Crivellin:2020lzu,Pich:2013lsa}. However, $C_{\phi \ell }^{(3)ij}$ also affects $Z$ couplings to leptons and neutrinos, which enter the global EW fit.

\section{Models addressing the CAA}

The following NP models (see Ref.~\cite{deBlas:2017xtg} for a complete categorization of tree-level extensions of the SM) give contributions to the effective operators discussed in the last subsection, which can explain the CAA.

\subsection{$W^\prime$ boson}

A $W^\prime$ boson with couplings to left-handed fermions (i.e.~the charged component of an $SU(2)_L$ triplet) affects the CAA via:
\begin{itemize}		\vspace{-1mm}	
	\item Modified $W\mu\nu$ coupling generated by the mixing with the SM $W$, encoded in $C^{(3)}_{\phi\ell}$. This leads to limited effects in the CAA due to the stringent constraints from $Z$ decays~\cite{Crivellin:2020ebi,Kirk:2020wdk}.		\vspace{-3mm}	
	\item Tree-level effects in  $\mu\to e\nu\nu$ via $C_{\ell \ell}^{(3)}$.	\vspace{-1mm}	
	\item Tree-level effects in $d\to u e\nu$ via $C_{\ell q}^{(3)}$ if the $W^\prime$ couples to quarks and leptons~\cite{Capdevila:2020rrl}. 		\vspace{-1mm}	
	\item Modified $W-u-d$ coupling generated by the mixing with the SM $W$ encoded in $C^{(3)}_{\phi q}$.
	\vspace{-1mm}	
\end{itemize}

\subsection{Vector-like Leptons}  

Vector-like leptons (VLL), such as right-handed neutrinos~\cite{Lee:1977tib}, affect $W-\ell-\nu$ coupling via their mixing with SM leptons and are EW symmetry breaking. There are 5 representations of vector-like leptons that can couple to SM leptons and the Higgs and mix with the former after EW symmetry breaking.  However, a single representation alone cannot explain the CAA and result simultaneously in a good EW fit~\cite{Crivellin:2020ebi,Kirk:2020wdk}. 

\subsection{Vector-like Quarks} 
Similar to vector-like leptons, vector-like quarks mix with the SM ones after EW symmetry breaking an lead to modified $W-u-d$ couplings. Two vector-like quarks are found of being capable of accounting for the CAA~\cite{Belfatto:2021jhf}.

\subsection{Singly charged $SU(2)_L$ singlet scalar}

Being an $SU(2)_L \times SU(3)_C$ singlet with hypercharge +1, only Yukawa-type interactions with leptons are allowed. Due to hermicity of the Lagrangian it has anti-symmetric (i.e. off-diagonal) couplings it results in an effect in muon decay with the right sign such that in can explain the CAA via an shift in the Fermi constant~\cite{Crivellin:2020klg,Crivellin:2020oup,Felkl:2021qdn,Marzocca:2021azj}. 

\subsection{$SU(2)_L$ Neutral Vector Boson ($Z^\prime$):}
A $Z^\prime$ boson which is an $SU(2)_L$ singlet only interferes with the SM amplitudes for $\mu\to e\nu\nu$ if it has flavour violating couplings to leptons. In this case a necessarily constructive effect is generated that can explain the CAA if other couplings are so small that the bounds from charged LFV are respected~\cite{Buras:2021btx}.

\subsection{Leptoquark}	

Even though LQs can generate tree-level effects in beta decays, in this case, low energy parity violation or kaon decays, are, in general, more constraining and prevent a solution of the CAA~\cite{Crivellin:2021egp,Crivellin:2021bkd}.

\section{Conclusions and outlook}

The Cabibbo angle, parametrizing the mixing among the first two quark generations can be measured in kaon, tau, pion, $D$ and beta decays. However, the determination from super-allowed beta decays disagrees with the other measurements, leading to an $\approx 3\,\sigma$ deficit in first row CKM unitarity, known as the CAA.

Since for extracting $V_{ud}$ from (super-allowed) beta decays knowledge of the Fermi constant is necessary. Even though the the Fermi constant is very precisely measured in muon decay, it is not necessarily free of new physics. THerefore, there are for ways how one can explain the CAA in terms of NP via effective operators: 	a) 4-fermion operators affecting beta decays ($u\to d e\nu$), b) operators generating modified $W$--$u$--$d$ coupling, c) 4-fermion operators affecting muon decays ($\mu\to e\nu\nu$) and
d) operators generating modified $W$--$\mu$--$\nu$ coupling. In terms of (simplified) BSM models, these operator can be generated by a) $W^\prime$, leptoquarks, b) $W^\prime$ (mixing with the SM $W$), vector-like quarks, c) $W^\prime$, singly charged scalar, $Z^\prime$ (with lepton flavour violating couplings), d) vector-like leptons, $W^\prime$ (mixing with the SM $W$). Note if only muon decay receives a BSM contribution, the CAA can be explained, but at the expense of a enhanced tension within the global EW fit.

\section*{Acknowledgements}

I thank the organisers for the invitation to Moriond and for assembling an extremely interesting program. Support via a Professorship Grant (PP00P2\_176884) of the Swiss National Science Foundation is gratefully acknowledged.

\end{document}